\documentstyle[12pt, epsf]{article}

\newcommand{\eq}[1]{Eq.~(\ref{#1})}

\def\bea{\begin{eqnarray}}
\def\eea{\end{eqnarray}}

\def\be{\begin{equation}}
\def\ee{\end{equation}}

\begin{document}
  \begin{flushright}
    SPIN-2001/17
  \end{flushright}
  \begin{center}
    {\Large \bf A note on the M5 brane anomaly.}\\[12pt]
    {\large A.Boyarsky\footnote{On leave of absence from Bogoliubov ITP, Kiev,
    Ukraine; e-mail:\ boyarsky@phys.uu.nl}\\ \emph{ Spinoza Institute, Leuvenlaan 4, 3584 CE, Utrecht,
  the~Netherlands}\\[6pt]
      B.Kulik\footnote{e-mail:\ bkulik@insti.physics.sunysb.edu}\\ \emph{C.N. Yang Institute for Theoretical Physics\\ SUNY at Stony Brook, NY 11794-3840, USA}}
  \end{center}
  
\begin{abstract}
The problem of the M5 brane anomaly cancellation is addressed. We
reformulate FHMM construction \cite{FHMM} making explicit the 
relation with the M5 brane SUGRA solution. We suggest another
solution to the magnetic coupling equation which doesn't need anomalous
$SO(5)$ variation of the 3-form potential and coincides with
the SUGRA solution outside smoothed out core of the magnetic source.
Chern-Simons term evaluated on this solution generates
the same anomaly inflow as achieved by FHMM.
\end{abstract}

The potential anomaly of the normal bundle in the presence of the
M5 brane has three contributions. The first one comes from a chiral theory of 
zero-modes on the brane world volume, second comes from the Chern-Simons
coupling $\int G_4 \wedge I_7$ , where $G_4$ is the 4-form field
strength and $I_7$ is a gravitational Chern-Simons.
The problem of finding the third contribution that cancels the previous
two is addressed in \cite{FHMM} in the following way. 

Instead of considering 11-d SUGRA in the background of the 11-d M5 brane
solution the other theory is studied. It's 11-d SUGRA with 
a 4-form field-strength satisfying modified Bianchi identity corresponding
to the (singular) magnetic coupling to a 6-dimensional 
submanifold (which represents the 5-brane in this picture). 

To have a well-defined delta-function in the magnetic coupling equation
the source is  smoothed out and
the 4-form G satisfies 
\bea 
d G = d \rho(r) \wedge e_4 /2~, \label{HdH}
\eea
 where $\rho$ is a bump function. It equals to $-1$ on the brane
($r=0$) and to $0$ far away from the brane, $e_4$ is an angular form
on the $SO(5)$ normal bundle. It is closed ($d e_4 = 0$) and gauge
invariant under $SO(5)$ transformations of the normal bundle. Locally
$e_4 = d e_3$ .
The tubular neighborhood of the brane
is removed and the resulting effective action in the bulk is 
\bea {\cal L} =
\lim_{\epsilon \rightarrow 0} \int_{M_{11} - D_\epsilon(W_6)} {\cal
  L}_{SUGRA}~. \label{HL} 
\eea 

In \cite{FHMM} it is argued that a general solution to \eq{HdH} is
\bea
G_4 = dC_3 + A \rho e_4 /2 - B d \rho \wedge e_3/2~. \label{HSH}
\eea

The requirement for $G$ to be regular at the origin
gives $A=0$ and $B=1$ (since $e_4$ is not well-defined at the origin,
but $d\rho=0$ at $r=0$). Thus
\bea
G_4 = dC_3 -  d \rho \wedge e_3/2~. \label{Hres}
\eea

Since $G$ should be gauge invariant under $SO(5)$
transformations but $e_3$ is not ( $\delta e_3 = d e^{(1)}_2$ )
$C$ has a gauge variation
\bea
\delta C = - d\rho \wedge e^{(1)}_2 /2~.
\eea

This leads to an anomalous variation of the modified Chern-Simons term in 
the action \eq{HL} and produces the necessary anomaly inflow from the bulk.
Chern-Simons should be modified since the relation between $G_4$ and
$d C_3$ has changed.

To maintain $d S_{CS} = ({\rm a~closed~form})^3$
Chern-Simons can be built out of
\bea
G_4 - \rho e_4/2 = d (C_3 - \rho e_3/2)~. \label{CSblock}
\eea

This is the choice of \cite{FHMM}. Thus the new Chern-Simons term
is
\bea
S_{CS} = \int_{M_{11} - D_\epsilon(W_6)} 
( C_3 - \rho e_3/2 ) \wedge d ( C_3 - \rho e_3/2 ) \wedge 
d ( C_3 - \rho e_3/2 )~.
\eea

In this approach the 5-brane is considered as a magnetic source 
for the 3-form and not as a solution to $D=11$ SUGRA.
``..the very important question of the relation of this approach to that 
based on a direct study of solutions to supergravity'' \cite{FHMM}
is left for the future.

Let us compare  \eq{Hres} with the background M5 brane solution \cite{M5}
\footnote{
By ${\bar \varepsilon}_{npqrs}$ we denote a flat 5-d anti-symmetric symbol.
The 11-d index $\mu$ is split into $(M,m)$, where $M$ is in the direction of 
the brane world volume and $m$ is in the direction transverse to the brane.
} 
\bea
d s^2 = \Delta^{-1/3} \eta_{MN} d x^M d x^N + 
\Delta^{2/3} \delta_{mn} d y^m d y^n \label{M5metric} ,\\
G = \frac{1}{4!}~\delta^{mn} \partial_m \Delta {\bar \varepsilon}_{npqrs}
d y^p \wedge d y^q \wedge d y^r \wedge d y^s , \label{M5form}\\
\Delta = 1 + \bigg(\frac{R}{r}\bigg)^3 ~,~~~~ r = \sqrt{\delta_{mn} y^m y^n}~.
\eea

Here $\Delta$ is a harmonic function such that $\Box \Delta = \delta(r)$
that is $(r^4  \Delta')' = \delta(r)$ 
($'$ is a derivative with respect to $r$).
The 4-form $G$ in the solution can be rewritten in the form 
\bea
G =  f(r) e_4/2 \label{OH}
\eea
(it is shown in the Appendix). 
Where
$f(r) = 1$ for $r>0$ and 
it jumps at $r=0$ since $f(r)'= \delta(r)$.

Therefore the background solution of SUGRA for 5-brane satisfies the equation
\bea
d G= \delta(r) e_4/2~.
\eea

Let us regularize the delta function in the spirit of \cite{FHMM} i.e.
find a corresponding solution of the magnetic coupling \eq{HdH}.

Away from the brane $G$ has the form $e_4/2$ for a magnetic brane
with a charge 1. We see that $dC$ in \eq{HSH} should be equal to $e_4$ for the 
background solution. Thus $C$ exists only locally and is equal to $e_3$
(up to a closed form). Therefore the general solution of magnetic coupling 
equation  $d G = d\rho\wedge e_4/2$  can better be written in the form 
\bea
G = d\tilde C + A \rho \wedge e_4/2 - B d\rho\wedge e_3/2 + ({\rm closed~4-form}) ~,
\eea
where $\tilde C$ is already globally defined.
To satisfy the asymptotic behavior of the background solution we have to take 
this closed 4-form to be $e_4$ (and $d\tilde C=0$ for background). 
Then if we set $A=1$ and $B=0$ we have
\bea
G = d\tilde C + (\rho+1) e_4/2~. \label{Greg}
\eea

It is still regular at $r=0$ since $\rho(0) = -1$.
In this case $C$ doesn't have any anomalous variation under
$SO(5)$ transformations since $e_4$ is gauge invariant.

Therefore, we just showed that there is 
a solution of the magnetic coupling equation 
which coincides with the classical solution to SUGRA outside a tubular 
neighborhood of the M5-brane (where $\rho=0$)
and is smooth near the brane. Let's see what anomaly inflow such
a choice of the solution leads to.
Evaluated on the class of solutions
\bea
G = d {\tilde C} + e_4/2 \label{Gnonreg}
\eea
the gauge variation of the Chern-Simons has necessary surface term. Indeed,
the potential for \eq{Gnonreg} can be defined locally as
$C =  {\tilde C} + e_3/2 $. Thus the variation of the Chern-Simons
is
\bea
\delta S_{CS} = \delta \int_{M_{11} - D_\epsilon(W_6)} 
({\tilde C} + e_3/2 )\wedge 
d ({\tilde C} + e_3/2 ) \wedge 
d ({\tilde C} + e_3/2 ) = \label{NonregCS}\\
\int_{M_{11} - D_\epsilon(W_6)} d e_2^{(1)}/2 \wedge 
(d {\tilde C} + e_4/2) \wedge (d {\tilde C} + e_4/2 )~.
\eea

Integrating by parts and taking the limit and using the fact that
$\tilde C$ is smooth near the brane we obtain
\bea
\delta S_{CS} = \int_{S_\epsilon(W_6)} e_2^{(1)}/2 \wedge e_4/2 \wedge e_4/2~.
\eea

This is in accord with \cite{FHMM}. The anomaly inflow is generated without
any smoothing out of the magnetic source.

As for the case of the regularized solution of \eq{Greg} , 
the Chern-Simons should be modified since $G$ is not a closed form.
One can make the same choice as in \cite{FHMM} , namely ,
take a closed form
$G - \rho e_4/2$  instead (see \eq{CSblock}). In this case
$G - \rho e_4/2 = e_4/2 + d {\tilde C} $ and
the modified Chern-Simons reads
\bea
S_{CS} = \int_{M_{11} - D_\epsilon(W_6)} ({\tilde C} + e_3/2)
\wedge d ({\tilde C} + e_3/2) \wedge d ({\tilde C} + e_3/2)~.
\eea

This is the Chern-Simons (see \eq{NonregCS} )
evaluated on the class of M5 brane solution \eq{Gnonreg}

Note that the function $\rho$ introduced for the regularization of the
magnetic source doesn't enter in the modified Chern-Simons.


\paragraph{Acknowledgments:}
We would like to thank G.Bonelli, J.Harvey, R.Minasian, O.Ruchayskiy,
A.Tseytlin, P.Vanhove for fruitful discussions.
Work of A.B. was partially supported by CRDF grant UP1-2115.

\section*{Appendix: Derivation of $G = f(r) e_4/2$.}

The value of $G$ in the solitonic solution of \eq{M5form} can be written
as
\bea
G = \frac{1}{4!}
{\bar \varepsilon}_{\mu\nu\gamma\lambda\rho} 
\delta^{\rho\sigma} \partial_\sigma \Delta 
~dy^\mu\wedge dy^\nu \wedge dy^\gamma \wedge dy^\lambda
= {\bar *} n~,~~~~ n= d\Delta~.
\eea

We denote by ${\bar *}$ a 5-d (transverse) dual with 
respect to the flat metric.
We want to show that
a geometrical meaning of ${\bar *} n$ from the point of view 
of the embedding $W^6\subset M^{11}$ is
${\bar *} n =f(r) e_4 $, where $f(r) = 128 \pi^2 r^4 \Delta'$. In \cite{FHMM}
$e_4$ was 
\bea
 e_4 = \frac{1}{4!} \frac{1}{64 \pi^2} {\bar \varepsilon}_{klmnp}~
{\hat y}^k~ d {\hat y}^l \wedge d {\hat y}^m 
\wedge d {\hat y}^n \wedge d {\hat y}^p~,
\eea
where all indexes are contracted with the flat metric, $\mid y \mid = r$ ,
${\hat y}^k = y^k / r$  and
$d {\hat y}^k = d y^l / r ( \delta_l^k - \delta_{lm} y^m y^k / r^2 )$.
In the angular form $e_4$ there is in general some part that depends on the 
connection in the normal bundle.
In the background solution it is obviously zero (it can also be easily checked
by direct calculations).
To find a relation between the normal form $n$ and the angular form $e_4$
we re-write $e_4$ in the basis of $dy^m$ instead of
$d{\hat y}^m$ 
\bea
e_4 = \frac{1}{4!} \frac{1}{64 \pi^2} {\bar \varepsilon}_{klmnp}~ \frac{1}{r^5}
y^k~ d y^{l'} \wedge d y^{m'} \wedge d y^{n'} \wedge d y^{p'}
D_{l'}^l~D_{m'}^m~D_{n'}^n~D_{p'}^p~, \label{e4.dy}
\eea
where 
\bea
D_m^n &\equiv& \delta_m^n - y_m y^n / r^2 \\
d {\hat y}^m &=& \frac{1}{r} D_n^m d y^n~.
\eea

Since $y^{[m} y^{n]} = 0$ only the first term in each of $D_m^n$ contributes
in \eq{e4.dy} 
\bea
e_4 = \frac{1}{4!} \frac{1}{64 \pi^2} {\bar \varepsilon}_{klmnp}~ \frac{1}{r^5}
y^k~ d y^{l} \wedge d y^{m} \wedge d y^{n} \wedge d y^{p}~.
\eea

This expression should be compared to ${\bar *}n$ 
\bea
n = d \Delta ~,~~~~~
{\bar *}n = \frac{1}{4!} 
\delta^{rk} {\bar \varepsilon}_{rlmnp}~ \partial_k \Delta~ 
d y^{l} \wedge d y^{m} \wedge d y^{n} \wedge d y^{p}~.
\eea

Thus for $f(r) e_4/2$ to be equal to  ${\bar *}n $ the function $f(r)$ 
should be taken as 
\bea
f(r)~ y^m / (128 \pi^2 r^5) =
\delta^{mn} \partial_n \Delta(r) = \Delta'~y^m / r ~.
\eea

It implies 
\bea 
f(r) = 128 \pi^2 r^4 \Delta'(r) ~. \label{diff_rho}
\eea

This is what we intended to prove. Since $r^4 \Delta' =const$ for $r >0$
parameter $R$ in $\Delta$ can be chosen such that $f(r) = 1$.


\end{document}